\documentclass[12pt]{article}

\usepackage{a4}
\usepackage{a4wide}

\usepackage{amsmath}
\usepackage{amscd}
\usepackage{amsfonts}
\usepackage{amssymb}
\usepackage{mathrsfs}
\usepackage{fancybox} 
\usepackage{enumerate}

\usepackage{cite}
\usepackage{bm}
\usepackage{amscd}
\usepackage{graphicx}
\usepackage[dvips]{color}
\usepackage{epsfig}

\makeatletter

\@addtoreset{equation}{section}
\makeatother

\usepackage{amsthm} 

\def\tr{\mathrm{tr}}

\def\Re{\mathrm{Re}}

\def\half{{1\over2}}

\def\Ric{\mathrm{Ric}}

\def\={\stackrel{\bullet}{=}}

\def\({\left(}
\def\){\right)}
\def\[{\left[}
\def\]{\right]}

\def\mbf{\mathbf}

\def \be {\begin{equation}}
\def \ee {\end{equation}}
\def \beqa {\begin{eqnarray}}
\def \eeqa {\end{eqnarray}}
\def \beal#1 {\begin{align}#1\end{align}}
\def \bes#1 {\begin{equation}\begin{split}#1\end{split}\end{equation}}
\def \nn {\notag\\}

\def\bra#1{\langle #1 |}

\def\ket#1{|#1 \rangle}

\def\aver#1{\left\langle #1 \right\rangle}

\def\pmat#1{\begin{pmatrix}#1 \end{pmatrix} }

\makeatletter

\@addtoreset{equation}{section}
\makeatother

\begin{document}

\begin{titlepage}
\title{
\vspace{-2cm}
\begin{flushright}
\normalsize{ 
YITP-20-33 \\ 
}
\end{flushright}
       \vspace{1.5cm}
Holographic de Sitter Spacetime and Quantum Corrections to The Cosmological Constant
       \vspace{1.cm}
}
\author{
Shuichi Yokoyama\thanks{shuichi.yokoyama[at]yukawa.kyoto-u.ac.jp}\\[25pt] 
{\normalsize\it Center for Gravitational Physics,} \\
{\normalsize\it Yukawa Institute for Theoretical Physics, Kyoto University,}\\
{\normalsize\it Kitashirakawa Oiwake-cho, Sakyo-Ku, Kyoto, Japan}
\\[10pt]
}

\date{}

\maketitle

\thispagestyle{empty}


\begin{abstract}
\vspace{0.3cm}
\normalsize

A dynamical aspect of quantum gravity on de Sitter spacetime is investigated by holography or the dS/CFT correspondence.  
We show that de Sitter spacetime emerges from a free Sp($N$) vector model by complexifying the ghost fields and flowing them in parallel to the imaginary axis.  
We confirm that the emergence of de Sitter spacetime is ensured by conformal symmetry. We also compute the quantum corrections to the cosmological constant up to the next-to-leading order of the $1/N$ expansion in a proposed holographic approach. As a result the sub-leading corrections have the opposite sign to the classical value. This implies that a quantum gravity on de Sitter spacetime is perturbatively stable and quantum effects make the universe flatter and the cosmological constant smaller.
\end{abstract}
\end{titlepage}

\section{Introduction} 

Observation of high redshift supernova \cite{Riess:1998cb,Astier:2005qq} and cosmic microwave background \cite{Penzias:1965wn,Ade:2015xua} indicates the accelerating expansion of the current universe with non-zero vacuum energy. 
Vacuum energy in the universe, which may have dominated at early times to induce the inflationary expansion \cite{Sato:1980yn,Guth:1980zm}, is currently observed to be much smaller than the expected particle physics scale \cite{Weinberg:1988cp}. This mystery known as the cosmological constant problem has stimulated theoretical physicists to invent various new ideas and approaches \cite{Witten:2000zk,Carroll:2000fy,Padmanabhan:2002ji,Copeland:2006wr}, but still remains unsolved.

An innovative method to explore quantum gravity has been proposed known as holography, which realizes gravitational theory on a manifold by quantum field theory on its boundary \cite{tHooft:1993dmi,Susskind:1994vu}. 
\if(See also \cite{Bousso:2002ju}.) \fi
Holography for a gravitational theory on de Sitter spacetime  
has been proposed \cite{Strominger:2001pn}, in which dual quantum field theory at boundary is some conformal field theory (CFT) as in the case of the holography for AdS \cite{Maldacena:1997re}. 
In the dS/CFT correspondence, CFT is defined on Euclidean space and non-unitary. It has been claimed that the time emerges from a renormalization group flow at boundary \cite{Strominger:2001gp} as the emergence of an AdS radial coordinate from CFT \cite{Maldacena:1997re,Skenderis:2002wp}, and that a bulk wave function at late times is a generating function in CFT at boundary \cite{Maldacena:2002vr}. 
(See also \cite{Hull:1998vg,Nojiri:2001mf,Klemm:2001ea,Balasubramanian:2001nb,Petkou:2001nk,McInnes:2001zw,Bousso:2001mw\if, Balasubramanian:2002zh\fi} for other earlier studies.)

A concrete example of the dS/CFT correspondence has been put forward \cite{Anninos:2011ui}, which claims that a higher spin gravity theory on de Sitter spacetime corresponds to the singlet sector of a conformal Sp($N$) vector model at boundary in the large $N$ limit. This proposal may be viewed as the dS/CFT version of higher spin/vector model duality \cite{Klebanov:2002ja} including cases for interacting theories \cite{Vasiliev:2003ev,Vasiliev:1986td}. (See also \cite{Sezgin:2002rt,Argurio:2004lva,Giombi:2012ms}.)
The non-unitary nature of an Sp($N$) vector model comes from the fact that the system consists of Grassmann-valued scalar fields \cite{Kausch:1995py,LeClair:2006kb,LeClair:2007iy}.  
This proposal enables one to test the dS/CFT correspondence or higher spin/Sp($N$) vector model duality by explicit large $N$ calculation.  (See \cite{Das:2012dt,Anninos:2012ft,Anninos:2013rza} for related studies.) 

Recently a new approach for holography has been proposed \cite{Aoki:2015dla,Aoki:2016ohw,Aoki:2016env,Aoki:2017bru}, in which a holographic direction is emergent from a flow equation that coarse-grains operators non-locally \cite{Albanese:1987ds,Narayanan:2006rf}. 
This flow equation approach makes it possible to construct a bulk gravitational theory explicitly, and turns out not to be restricted to AdS nor to classical geometry \cite{Aoki:2017uce,Aoki:2018dmc,Aoki:2019bfb,Aoki:2019dim}.
The aim of this paper is to combine the proposals and techniques mentioned above to construct de Sitter spacetime and shed light on a dynamical aspect of quantum gravity on de Sitter spacetime from the holographic viewpoint.  

The rest of this paper is organized as follows. 
In section \ref{SpN}, we briefly study a free Sp($N$) vector model and discuss complexification of a general Sp($N$) vector model for later use. In section \ref{flow} we flow the complexified Sp($N$) ghost fields in parallel with the imaginary axis. We show that this smearing leads to the emergence of de Sitter spacetime. We give comments on a relation between the holographic metric and an information metric and on the emergence of de Sitter spacetime from the symmetry perspective. 
In section \ref{CC} we compute quantum corrections to the cosmological constant up to the next-to-leading order in the $1/N$ expansion. 
The last section is devoted to discussion. In appendix \ref{calculation} we give a derivation of a key formula in the main text. 

\section{${\rm Sp}(N)$ vector model} 
\label{SpN}

Following the proposal of the dS/CFT correspondence \cite{Anninos:2011ui} we start with a free ${\rm Sp}(N)$ vector model on a general $d$-dimensional Euclidean space, where $N$ is an even number and $d$ is greater than 2.\footnote{ 
In this paper we mainly focus on a free theory, in which case the restriction of the dimension is required for a general analysis. Two dimensional free theories need to be analyzed separately. 
}
The action is given by 
\beal{
S= \int d^dx {1\over 2} \Omega_{ab} \delta^{jk} \partial_j \chi^a \partial_k \chi^b,
}
where $\chi^a$ $(a=1,\cdots, N)$ denote the Grassmann-valued scalar fields and $\Omega_{ab}$ is the invariant tensor of Sp$(N)$ of symplectic form 
\be 
\Omega_{ab} = \pmat{0 & 1_{\frac N2\times \frac N2} \\ - 1_{\frac N2\times \frac N2} & 0}. 
\ee
From this action, the propagator of the scalar field is computed as
\beal{
& \aver{  \chi^c( x_1) \chi^b( x_2) } 
=\Omega^{cb}  { C \over (x_{12}^2)^{{d\over2}-1}},
\label{propagator}
}
where $C={ \Gamma({d\over2}-1) \over 4 \pi^{{d\over2}} }$, $\Omega^{cb}=\Omega_{bc}$, $x^k_{12}:= x_1^k - x_2^k$, $X^2= X^k X^k$.

Since Grassmann-valued scalar fields are ghosts, which disobey the spin-statistics theorem, Sp($N$) vector models with or without self-interaction are generally non-unitary quantum field theories. 
We define a quantum theory of Sp($N$) ghost fields by path integral. Practical evaluation of the path integral is done by the saddle point method, in which the integration contour should be deformed suitably in the complex plane. 
For convenience of later discussion, although the theory is originally defined on Euclidean space, 
we complexify both the base space coordinates and the ghost fields so that the contours of the path integral can be deformed in complex domains in a certain way \cite{Parisi:1984cs,Klauder:1983sp}.%
\footnote{ 
It is known that the complex Langevin method does not always give a correct prescription to deform integration contour in a convergent manner \cite{Aarts:2009uq,Aarts:2011ax}. This method can be improved by the Lefschetz thimble method \cite{Witten:2010cx,Cristoforetti:2012su}, which determine correct thimbles for the contour to be deformed for convergence. See also \cite{Fujii:2013sra,Kanazawa:2014qma,Fujii:2015bua,Nagata:2016vkn,Fukuma:2017fjq}. 
}
We denote a complex coordinate corresponding to a real coordinate by its bold letter and its imaginary part by the same letter with check: 
\be 
x^k \to \bm x^k=x^k+ i\check x^k.
\ee
The ghost fields are notationally intact. 
Accordingly the action and observables are extended in a holomorphic fashion. 
Then the two point function given in \eqref{propagator} is now extended for the complexified ghost fields to 
\beal{
\aver{  \chi^c(\bm x_1) \chi^b(\bm x_2) } 
=\Omega^{cb}  { C \over (\bm x_{12}^2)^{{d\over2}-1}} .
\label{2ptcomplex}
}
Note that the holomorphically extended theory enjoys the complexified conformal symmetry
\beal{ 
\delta^{\rm conf}_{\delta \bm x}\chi  =& - {\delta \bm x}^k\partial_k\chi  -  {d-2 \over 2d}  \partial_k \delta \bm x^k\chi
\label{conformalhol}
}
with $\delta \bm x^k = a^k + \omega^k{}_j \bm x^j + \lambda \bm x^k + b^k \bm x^2 -2 \bm x^k (b_j \bm x^j)$. This determines the coordinate dependence of the two point function \eqref{2ptcomplex}.

\section{Imaginary flow and emergent conformal time} 
\label{PIF}

Let us consider the construction of de Sitter spacetime from Sp($N$) vector models formulated in the previous section by the flow equation approach. 
To this end the first thing to do is to choose an appropriate flow equation for Sp($N$) ghost fields. 
We know that the same flow equation as used previously leads to emergence of Euclidean AdS from a general result \cite{Aoki:2017bru}, so we need to change a flow equation suitably. 
One naive guess may be to flip the sign of the flow parameter because the de Sitter metric is obtained by the analytic continuation of Euclidean AdS:
\be 
-{\partial \over \partial s}\chi^a(x;s) \overset?= \partial^2 \chi^a(x;s), 
\ee
where $s$ is a positive number corresponding to a flow parameter, and $\partial^2=\partial_k\partial_k$. 
However this is not allowed since the resulting flowed operator is no more well-defined to give a finite correlation function. 

Then what is a correct flow equation for de Sitter spacetime to emerge? Our answer is to complexify the theory as explained in the previous section and flow the ghost operators in parallel to the imaginary axis:%
\footnote{
This is very similar to construction of bulk local operators from CFT known as the HKLL construction \cite{Hamilton:2006az} in the regard to require the complexification of the boundary coordinates for smeared operators not to be divergent.
Differences of two ways of smearing are not only in the smearing functional but also in the smearing region: In the HKLL case, the smearing domain is restricted to a causal wedge diamond, while in the imaginary flow here operators are smeared fully in the imaginary part. The relation between the two methods is not clarified yet.
The author would like to thank Tatsuma Nishioka for valuable discussion on this point. 
}
\be 
{\partial \over \partial \xi}\chi^a(\bm x;\xi) = \check\partial^2 \chi^a(\bm x;\xi), 
\label{flow}
\ee
where $\check\partial_k:= \partial_{\check x^k}$. We refer to this flow as an imaginary flow and the conventional one as a real flow. 
This is solved by the convolution with a gaussian in terms of the imaginary coordinates: 
\beal{  
\chi^a(\bm x;\xi) 
=\int_{\mbf R^d} d^d\check x' \rho(\check x,\check x';\xi) \chi^a(x+i\check x') , ~~~ 
\rho(\check x,\check x';\xi)
=& {e^{ - {(\check x-\check x')^2 \over 4\xi} } \over (4\pi\xi)^{d\over2}}.
\label{flowedop}
}
We claim that the smeared operators by the imaginary flow are well-behaved for an arbitrary real coordinates $x^k$ and any positive flow parameter $\xi$ if the smeared operators are inserted with sufficient separation with respect to the imaginary coordinates. Let us see this by computing the two point function. The result is
\be 
\aver{\chi^a(\bm x_1;\xi_1) \chi^b(\bm x_2;\xi_2) } 
={\Omega^{ab}\over i^{d-2} (\xi_1+\xi_2)^{{d-2\over2}}} F( {(\check x_{12} - ix_{12})^2 \over \xi_1+\xi_2}), 
\label{2ptflowchi}
\ee
where $F(s)$ is defined by 
\beal{
F(s)
=&{ 1 \over (4\pi)^{{d\over2}}}\int_0^1 dv  e^{{-sv \over 4} } v^{{d\over2} -2}. 
\label{F}
}
We derive \eqref{2ptflowchi} in appendix \ref{calculation}.
It is clearly seen that the two point function reproduces the original correlation function in the limit $\xi_1,\xi_2\to0$ if we consider $|\check x_{12}^k| > |x_{12}^k|$.

Notice that the two point function \eqref{2ptflowchi} is finite in the contact limit $(\bm x_2^k,\xi_2)\to (\bm x_1^k,\xi_1)$.%
\footnote{ 
To be consistent with the separation of inserted smeared operators described above, the contact limit is taken first for the original real coordinates and the flow parameter $(x_2^k,\xi_2)\to (x^k_1,\xi_1)$ and subsequently for the imaginary coordinates $\check x_2^k \to \check x_1^k$. 
}
This property enables us to define a metric operator using a normalized field \cite{Aoki:2015dla,Aoki:2016ohw,Aoki:2016env}
\be 
\sigma^a(\bm x;\xi) = { \chi^a(\bm x;\xi)  \over (\aver{\chi^c(\bm x;\xi)\Omega_{cb}\chi^b(\bm x;\xi)})^\half }.
\label{normalizedfield}
\ee
\footnote{The normalization factor is computed as $
\aver{\chi^c(\bm x;\xi)\Omega_{cb}\chi^b(\bm x;\xi)}
={4N \xi \over i^{d} (8\pi\xi)^{{d\over2}}  (d-2)}.$}
For later use, we write down the two point function of the normalized field
\beal{ 
\aver{\sigma^a(\bm x_1;\xi_1) \sigma^b(\bm x_2;\xi_2) } 
=-{ \Omega^{ab} \over N} ({\sqrt{4\xi_1\xi_2}\over \xi_1+\xi_2})^{d-2\over2} { F({(\check x_{12} - ix_{12})^2  \over \xi_1+\xi_2}) \over F(0)}.
\label{2ptnormalizedfield}
}
Using this normalized field, we define the metric operator by contracting the Sp($N$) indices as 
\beal{ 
\hat g_{\mu\nu}(\bm x;\xi)
=&- \partial_\mu \sigma^a(\bm x;\xi) \Omega_{ab}\partial_\nu \sigma^b(\bm x;\xi), 
\label{metricop}
} 
where $\partial_\mu:= \partial_{\bm X^\mu}$ with $\bm X^\mu:=(\bm x^k,\eta), \eta \propto \sqrt\xi$, so that $\eta$ has one length dimension. The overall minus sign is a convention. 
Then the metric in the holographic space is defined by the vacuum expectation value (VEV) of this metric operator with the imaginary coordinates set to zero:%
\footnote{In most practical cases, the system enjoys the translational symmetry. In this case the holographic metric is independent of the $\bm x^k$ coordinates, so that it does not need to set the imaginary coordinates to zero.} 
\be
ds^2 =\lim_{\check x^k\to0} \aver{ \hat g_{\mu\nu}(\bm x;\xi) } dX^\mu dX^\nu.
\label{holmetric}
\ee
The VEV of the metric operator is computed as 
\beal{ 
\aver{\hat g_{\xi\xi}(\bm x;\xi)} =&- {d-2 \over 8\xi^2 }, ~~ 
\aver{\hat g_{ij}(\bm x;\xi)} 
= -{\delta_{ij} \over \xi} {F'(0) \over F(0)}, 
}
and other components vanish. 
Plugging this back to \eqref{holmetric} gives  
\beal{
ds^2 
=& (- {d-2 \over 8\xi^2 }d\xi^2 - {F'(0) \over F(0)}{1\over \xi}  (dx^k)^2 ).
}
Using \eqref{F} we find ${F'(0) \over F(0)} = -{d-2 \over 4d}$. Therefore setting $\eta =\sqrt{ 2d\xi }$
we find that the holographic metric becomes 
\beal{
ds^2=& {d-2\over 2} \({- d\eta^2 + (dx^k)^2 \over \eta^2} \) . 
\label{dSmetric}
}
This is nothing but the metric of de Sitter spacetime with the radius $\sqrt{d-2\over2}$ and the (conformal) time $\eta$. 
As a result, de Sitter spacetime emerges from an Sp($N$) vector model via the imaginary flow. 

\subsection{Relation to information metric} 
\label{IM} 

In the previous study using a real flow, the holographic metric for Euclidean CFT, which generates an Euclidean AdS metric, can be regarded as an information metric \cite{Aoki:2017bru}. In the current case using the imaginary flow, we show that the holographic metric becomes de Sitter one. This means that it is not positive definite any more. Then can it still be regarded as the information metric?

To answer this, let us trace the argument in \cite{Aoki:2017bru} and apply it to the present case. For this purpose we recall the information metric or the Bures metric in an arbitrary quantum mechanical system, which is defined by
\be
D(\rho,\rho+d\rho)^2 = \frac{1}{2}{\rm tr} (d\rho\, G),
\ee
where $\rho$ and $\rho+d\rho$ are density matrices in the system, and $G$ is an Hermitian 1-form operator to satisfy an equation 
\be 
\rho\, G + G\, \rho = d\rho.
\label{informationG}
\ee 
It is known that this gives a well-defined metric for the space of density matrices in the system. 

Let us consider an Sp($N$) vector model and flow the Sp($N$) vector matter by the imaginary one as in \eqref{flow}.\footnote{The discussion in this subsection can be applied to any Sp($N$) vector models with or without self-interaction.} 
Then the quantity 
\be 
\rho_{\bm X} =\sigma^a(\bm x;\xi)\ket{0} \Omega_{ab}\bra{0}\sigma^b(\bm x;\xi),
\ee
where $\ket0$ is the vacuum of the system,
becomes a density matrix thanks to the normalization condition \eqref{normalizedfield}:
\be 
\tr \rho_{\bm X} = \aver{\sigma^a(\bm x;\xi)\Omega_{ab} \sigma^b(\bm x;\xi)} = 1. 
\ee
It can be also shown that $ \rho_{\bm X}^2 =-{1\over N} \rho_{\bm X}$.
Therefore if we define $G_{\bm X} = -N d \rho_{\bm X}$, then this is a Hermitian 1-form operator to satisfy $\rho_{\bm X}\, G_{\bm X} + G_{\bm X}\, \rho_{\bm X} = d\rho_{\bm X}.$
In this setup, the information metric is computed as 
\beal{
D(\rho_{\bm X},\rho_{\bm X}+d\rho_{\bm X})^2 =- \aver{g_{\mu\nu}(\bm x;\xi) }d\bm X^\mu d\bm X^\nu, 
}
where we used $d\rho_{\bm X}=d\bm X^\mu \partial_\mu \rho_{\bm X}$. 
As a result, the information metric agrees with the holographic metric given by \eqref{holmetric} up to a signature after sending the imaginary coordinates to zero. The holographic metric can be interpreted as an information metric even in a non-positive definite case.

\subsection{Isometry from conformal symmetry} 
\label{symmetry}

It can be shown that the emergence of de Sitter spacetime from a free Sp($N$) vector model via the flow equation approach presented above is guaranteed by conformal symmetry at boundary as in the AdS case \cite{Aoki:2017bru}.\footnote{The symmetry argument in this subsection is not restricted to a free case, and can be extended to an arbitrary CFT by replacing the ghost field $\chi$ and its conformal dimension ${d-2\over2}$ with an scalar primary operator of a general dimension as in the AdS case. 
}  

Let us see this briefly. To this end we compute the complexified conformal transformation \eqref{conformalhol} for the normalized field given by \eqref{normalizedfield}:
\beal{  
\delta^{\rm conf}\sigma^a
=&- (\xi \check\partial^2 \delta \bm x^k + 2\xi^2  \check\partial^j \check\partial^k \delta \bm x^k \check\partial_j \check\partial_k   + 2\xi \check\partial^j \delta \bm x^k \check\partial_j + \delta \bm x^k) \partial_k \sigma^a - {d-2 \over 2d}(2\xi \check\partial^j \partial_k {\delta \bm x}^k \check\partial_j + \partial_k {\delta \bm x}^k) \sigma^a. \notag
}
Using the explicit form of $\delta\bm x^k$ and the flow equation \eqref{flow}, $\partial_\xi \sigma^a =({d-2 \over 4\xi } + \check\partial^2) \sigma^a$, we rewrite this as 
\beal{  
\delta^{\rm conf}\sigma^a
=&- 2 {\xi}\lambda \partial_{\xi}  \sigma^a+ 4{\xi} (b_k\bm x^k) \partial_{\xi}  \sigma^a - 4{\xi}^2 b^j \partial_j (\partial_{\xi} + {d-2  \over 4{\xi}}) \sigma^a  - (\delta \bm x^k-2{\xi} (d-2) b^k  )\partial_k \sigma^a \notag.
}
There exists a higher derivative term for a special conformal transformation. Thus it cannot be regarded as an infinitesimal diffeomorphism in this form, so we separate into one derivative terms and higher derivative ones, $\delta^{{\rm conf}} \sigma^a  = \delta^{{\rm diff}} \sigma^a + \delta^{{\rm extra}} \sigma^a $, where 
\beal{ 
\delta^{{\rm diff}} \sigma^a =&  (- 2 {\xi}\lambda + 4{\xi} (b_k\bm x^k) ) \partial_{\xi}  \sigma^a  - (\delta \bm x^k-2{\xi} (d-2+a) b^k  )\partial_k \sigma^a , \\ 
\delta^{{\rm extra}} \sigma^a
=&-4{\xi}^2 b^j \partial_j (\partial_{\xi} + {d-2 +2a \over 4{\xi}})  \sigma^a .
\label{separation}
}
Here $a$ is a constant to be determined to satisfy $\aver{\delta^{{\rm diff}} \hat g_{\mu\nu}(\bm x;\xi)}=0$. This is equivalent to the condition
\be 
\lim_{\bm X_2\to \bm X_1}\partial_{\bm X^\mu_1} \partial_{\bm X^\nu_2} \aver{ \delta^{{\rm extra}}\sigma^a(\bm x_1;\xi_1)\Omega_{ab} \sigma^b(\bm x_2;\xi_2)  +\sigma^a(\bm x_1;\xi_1)\Omega_{ab} \delta^{{\rm extra}}\sigma^b(\bm x_2;\xi_2)  } =  0,
\label{condition}
\ee
since a correlation function is invariant any conformal transformation $\aver{\delta^{{\rm conf}} \hat g_{\mu\nu}(\bm x;\xi)}=0$. 
Employing \eqref{2ptnormalizedfield}
we can compute the average in the right hand side of \eqref{condition} as   
\beal{
&\aver{\delta^{{\rm extra}}\sigma^a(\bm x_1;\xi_1)\Omega_{ab} \sigma^b(\bm x_2;\xi_2)  + \sigma^a(\bm x_1;\xi_1)\Omega_{ab}\delta^{{\rm extra}}\sigma^b(\bm x_2;\xi_2)  } \nn
=&4{ (2\sqrt{\xi_1\xi_2})^{d-2\over2} \over  (\xi_1+\xi_2)^{\frac d2+2}}(\xi_1^2  -\xi_2^2 ) {2b_k \bm x_{12}^k }{\bm x_{12}^2 } {F''({-\bm x_{12}^2 \over \xi_1+\xi_2}) \over F(0)} -(4-2a)(\xi_1  -\xi_2 ){ (2\sqrt{\xi_1\xi_2})^{d-2\over2} \over  (\xi_1+\xi_2)^{d\over2}} {2b_k \bm x_{12}^k} {F'({-\bm x_{12}^2 \over \xi_1+\xi_2}) \over F(0)} \notag.
}
This shows that the condition \eqref{condition} is satisfied if and only if $a=2$. 
In this case, the holographic metric \eqref{holmetric} is invariant under the following infinitesimal diffeomorphism  
\bes{
\bar\delta x^k =&\delta x^k -2\xi d b^k, \quad 
\bar\delta {\xi} = 2\lambda \xi - 4\xi b_j x^j, 
} 
where we set the imaginary coordinates to zero. 
By using the conformal time $\eta$ this can be rewritten as 
\beal{
\bar\delta x^k =&\delta x^k - \eta ^2 b^k   , \quad 
\bar\delta \eta  = \lambda \eta  - 2\eta b_j x^j. 
\label{isomDS}
}
This is nothing but the isometry transformation of de Sitter spacetime. Since the de Sitter spacetime is maximally symmetric, the isometry restricts the coordinate dependence of the metric completely as in \eqref{dSmetric}.

\section{Quantum corrections to the cosmological constant} 
\label{CC}

In the flow equation approach, holographic computation of quantum corrections to a cosmological constant in the bulk is clarified, and the explicit calculations were carried out up to the next-to-leading order for a dual bulk theory to a free O($N$) vector model \cite{Aoki:2018dmc}. Following this procedure we compute the one loop corrections to the cosmological constant for a dual bulk theory to a free Sp($N$) vector model. 

To this end, it is convenient to introduce pre-geometric operators, which are given from a usual geometric object known in differential geometry by replacing the metric tensor with the metric operator. 
For example, the Levi-Civita connection operator and curvature operators are defined by  
\beal{
\hat\Gamma^{ \mu }_{\rho \nu }(\bm x;\xi) =& \half \hat g^{ \mu  \sigma }(\bm x;\xi) (\hat g_{ \sigma \{ \nu , \rho \}}(\bm x;\xi) -\hat g_{ \nu  \rho , \sigma }(\bm x;\xi) ), \nn
\hat R_{ \rho  \sigma }{}^{ \mu }{}_{ \nu }(\bm x;\xi)=&\partial_{[ \rho }\hat \Gamma^{ \mu }_{ \sigma ] \nu }(\bm x;\xi)+\hat\Gamma_{[ \rho\lambda }^{ \mu }(\bm x;\xi)\hat\Gamma_{ \sigma ]\nu}^{ \lambda }(\bm x;\xi), \nn
\hat \Ric_{ \sigma  \nu }(\bm x;\xi)=&\hat R_{ \mu  \sigma }{}^{ \mu }{}_{ \nu }(\bm x;\xi), \nn
\hat R(\bm x;\xi)=&\hat g^{ \sigma  \nu }(\bm x;\xi)\hat R_{ \sigma  \nu }(\bm x;\xi), 
}
where $\hat g^{ \mu\sigma}(\bm x;\xi) $ is the inverse of the metric operator $\hat g_{ \mu\sigma}(\bm x;\xi) $, which exists at least perturbatively, and $X_{\{x,y\}}:= X_{x,y} + X_{y,x}$, $X_{[x,y]}:= X_{x,y} -X_{y,x}$. First let us compute the VEV of the Einstein tensor operator up to the next-to-leading order of the $1/N$ expansion:
\be 
\hat G_{\mu\nu}(\bm x;\xi) = \hat R_{\mu\nu}(\bm x;\xi) - \half \hat g_{\mu\nu}(\bm x;\xi) \hat R(\bm x;\xi) .
\ee

In the leading order of the $1/N$ expansion, the VEVs of these operators are identical to the classical values of $d+1$ dimensional de Sitter spacetime.%
\footnote{Explicitly, $
\Gamma^{\mu}_{\nu\rho} 
={ g^{\mu\eta} g_{\nu\rho} - \delta^\mu_{\{\nu} \delta^\eta_{\rho\}} \over \eta}, ~ 
R_{ \rho  \sigma }{}^{ \mu }{}_{ \nu }={ \delta^\mu_{[\rho} g_{\sigma]\nu} \over L_{\rm dS}^2}, ~ 
\Ric_{\mu\nu}= d {g_{\mu\nu}\over L_{\rm dS}^2}, ~ 
R={d(d+1) \over L_{\rm dS}^2}. $
}
In particular, the VEV of the Einstein tensor is $
G_{\mu\nu}= - \Lambda g_{\mu\nu},$
where $\Lambda = {d(d-1) / (2 L_{\rm dS}^2})$ is the cosmological constant with $L_{\rm dS}$ the de Sitter radius classically given by $L_{\rm dS}=\sqrt{(d-2)/2}$. 

In the next-to-leading order, we need to perform the covariant perturbation for the metric operator. Since the general framework is fully developed in \cite{Aoki:2018dmc}, here we explain only the main results and skip the detail of lengthy computations. 
We perturb the metric operator as $\hat g_{\mu\nu} = g_{\mu\nu} + \hat h_{\mu\nu}$, where $g_{\mu\nu}=\aver{\hat g_{\mu\nu}}$, 
and expand the pre-geometric operators in terms of the fluctuation field $\hat h_{\mu\nu}$. Since the one point function of $\hat h_{\mu\nu}$ vanishes, the corrections starts from the quadratic order. 
The quadratic order of the Einstein tensor operator reads \cite{Aoki:2018dmc}
\beal{
\ddot G_{\mu\nu}
=&\ddot\Ric_{\mu\nu} -\half \hat h_{\mu\nu} \dot R -\half g_{\mu\nu} \ddot R, 
}
where 
\bes{
\dot R=&-\nabla^2 \hat h^{\nu}{}_{\nu}+\hat h^{\mu\nu}{}_{;\mu\nu} -\hat h^{\nu\sigma} \Ric_{\nu\sigma}, \\
\ddot\Ric_{\sigma\nu}=&\half [- \nabla_\mu \{ \hat h^{\mu\kappa}(\hat h_{\kappa\{\nu; \sigma\}}-\hat h_{\nu\sigma;\kappa})\} + \nabla_\sigma \{ \hat h^{\mu\kappa}(\hat h_{\kappa\{\nu; \mu\}}-\hat h_{\nu\mu;\kappa})\} ] \nn
& + {1\over4}[ \hat h^\mu{}_{\mu; \rho} (\hat h^\rho{}_{\{\nu; \sigma\}}-\hat h_{\nu\sigma}{}^{;\rho})- (\hat h^\mu{}_{\{\sigma; \rho\}}-\hat h_{\sigma\rho}{}^{;\mu}) (\hat h^\rho{}_{\{\nu; \mu\}}-\hat h_{\nu\mu}{}^{;\rho})], \\
\ddot R=& g^{\nu\sigma} \ddot\Ric_{\sigma\nu} +\half \hat h^{\nu\sigma}(\nabla^2 \hat h_{\sigma\nu}-2\hat h^{\mu}{}_{\sigma;\nu\mu}+\hat h^\mu{}_{\mu}{}_{;\nu\sigma}) + \hat h^{\rho\sigma}\hat h_{\sigma}{}^\nu \Ric_{\nu\rho}.  
}
Here $\nabla_\mu$ is the covariant derivative defined by the background metric $g_{\mu\nu}$ and we denote $X_{;\mu}=\nabla_\mu X$. 
The VEV of each term can be computed from two point functions of $\hat h_{\mu\nu}$ and its covariant derivatives.
The result is%
\footnote{ 
The following data, which is computed by mathematic, is used for the next-to-leading computation.
\beal{
&\aver{\hat h_{\mu\nu} \hat h_{\rho\sigma} }
=- {g_{\{\mu\rho}g_{\nu\}\sigma} \over N }, ~~
\aver{\hat h_{\mu\nu;\omega} \hat h_{\rho\sigma} }
=0,  \nn
&\aver{\hat h_{\mu\nu;\omega} \hat h_{\rho\sigma;\kappa} }
=-\aver{\hat h_{\mu\nu;\omega\kappa} \hat h_{\rho\sigma} }
\nn
=&\frac{i^{\#} (d-2)^2}{8 (d+2) N \eta^6}\big[ (d+1)^2 \{ 2 \delta_{\kappa\omega } \delta_{\mu\sigma } \delta_{\nu\rho }  +\delta_{\kappa\sigma } \delta_{\mu\omega } \delta_{\nu\rho } +2 \delta_{\kappa\omega } \delta_{\mu\rho } \delta_{\nu\sigma }  +\delta_{\kappa\rho } \delta_{\mu\omega } \delta_{\nu\sigma } +\delta_{\kappa\sigma } \delta_{\mu\rho } \delta_{\nu\omega } +\delta_{\kappa\rho } \delta_{\mu\sigma } \delta_{\nu\omega } \nn
&+\delta_{\kappa\nu } \delta_{\mu\sigma } \delta_{\rho\omega }  +\delta_{\kappa\mu } \delta_{\nu\sigma } \delta_{\rho\omega } +\delta_{\kappa\nu } \delta_{\mu\rho } \delta_{\sigma\omega } +\delta_{\kappa\mu } \delta_{\nu\rho } \delta_{\sigma\omega }  \} \nn
&+(d+1) \{-4 \delta_{\kappa\omega } \delta_{\mu\sigma } \delta_{\nu\rho }  -4 \delta_{\kappa\sigma } \delta_{\mu\omega } \delta_{\nu\rho }  -4 \delta_{\kappa\omega } \delta_{\mu\rho } \delta_{\nu\sigma }  -4 \delta_{\kappa\rho } \delta_{\mu\omega } \delta_{\nu\sigma }  -4 \delta_{\kappa\sigma } \delta_{\mu\rho } \delta_{\nu\omega }  -4 \delta_{\kappa\rho } \delta_{\mu\sigma } \delta_{\nu\omega } \nn
&+ 2 \delta_{\eta\nu } \delta_{\eta\omega } \delta_{\kappa\sigma } \delta_{\mu\rho }  +2 \delta_{\eta\rho } \delta_{\eta\omega } \delta_{\kappa\nu } \delta_{\mu\sigma }  +2 \delta_{\eta\nu } \delta_{\eta\omega } \delta_{\kappa\rho } \delta_{\mu\sigma }  +2 \delta_{\eta\nu } \delta_{\eta\rho } \delta_{\kappa\omega } \delta_{\mu\sigma }  +2 \delta_{\eta\mu } \delta_{\eta\omega } \delta_{\kappa\sigma } \delta_{\nu\rho } +2 \delta_{\eta\rho } \delta_{\eta\omega } \delta_{\kappa\mu } \delta_{\nu\sigma } \nn
&+2 \delta_{\eta\mu } \delta_{\eta\omega } \delta_{\kappa\rho } \delta_{\nu\sigma }  +2 \delta_{\eta\mu } \delta_{\eta\rho } \delta_{\kappa\omega } \delta_{\nu\sigma }-2 \delta_{\kappa\nu } \delta_{\mu\sigma } \delta_{\rho\omega }  -2 \delta_{\kappa\mu } \delta_{\nu\sigma } \delta_{\rho\omega }  -2 \delta_{\kappa\nu } \delta_{\mu\rho } \delta_{\sigma\omega }  -2 \delta_{\kappa\mu } \delta_{\nu\rho } \delta_{\sigma\omega } \} \nn
&-2 \delta_{\eta\nu } \delta_{\eta\omega } \delta_{\kappa\sigma } \delta_{\mu\rho }-2 \delta_{\eta\rho } \delta_{\eta\omega } \delta_{\kappa\nu } \delta_{\mu\sigma }-2 \delta_{\eta\nu } \delta_{\eta\omega } \delta_{\kappa\rho } \delta_{\mu\sigma } -2 \delta_{\eta\nu } \delta_{\eta\rho } \delta_{\kappa\omega } \delta_{\mu\sigma }-2 \delta_{\eta\mu } \delta_{\eta\omega } \delta_{\kappa\sigma } \delta_{\nu\rho }+2 \delta_{\kappa\omega } \delta_{\mu\sigma } \delta_{\nu\rho } \nn
&+2 \delta_{\kappa\omega } \delta_{\mu\rho } \delta_{\nu\sigma } -2 \delta_{\eta\rho } \delta_{\eta\omega } \delta_{\kappa\mu } \delta_{\nu\sigma }-2 \delta_{\eta\mu } \delta_{\eta\omega } \delta_{\kappa\rho } \delta_{\nu\sigma }-2 \delta_{\eta\mu } \delta_{\eta\rho } \delta_{\kappa\omega } \delta_{\nu\sigma } \nn
&-\delta_{\kappa\sigma } \delta_{\mu\omega } \delta_{\nu\rho }-\delta_{\kappa\rho } \delta_{\mu\omega } \delta_{\nu\sigma }-\delta_{\kappa\sigma } \delta_{\mu\rho } \delta_{\nu\omega }-\delta_{\kappa\rho } \delta_{\mu\sigma } \delta_{\nu\omega }+\delta_{\kappa\nu } \delta_{\mu\sigma } \delta_{\rho\omega }+\delta_{\kappa\mu } \delta_{\nu\sigma } \delta_{\rho\omega }+\delta_{\kappa\nu } \delta_{\mu\rho } \delta_{\sigma\omega }+\delta_{\kappa\mu } \delta_{\nu\rho } \delta_{\sigma\omega } \nn
&+2d\big\{ \delta_{\eta\sigma } \delta_{\kappa\omega } (\delta_{\eta\nu } \delta_{\mu\rho }+\delta_{\eta\mu } \delta_{\nu\rho})+\delta_{\eta\sigma }\delta_{\eta\omega } (\delta_{\kappa\nu } \delta_{\mu\rho }+\delta_{\kappa\mu } \delta_{\nu\rho })  \nn
&+ \delta_{\eta\kappa } (2 \delta_{\eta\omega } \delta_{\mu\sigma } \delta_{\nu\rho }+\delta_{\eta\sigma } \delta_{\mu\omega } \delta_{\nu\rho }+2 \delta_{\eta\omega } \delta_{\mu\rho } \delta_{\nu\sigma }+\delta_{\eta\rho } \delta_{\mu\omega } \delta_{\nu\sigma } +\delta_{\eta\sigma } \delta_{\mu\rho } \delta_{\nu\omega }+\delta_{\eta\rho } \delta_{\mu\sigma } \delta_{\nu\omega } ) \nn
&+\delta_{\eta\nu } (-3 \delta_{\eta\sigma } \delta_{\eta\omega } \delta_{\mu\rho }+\delta_{\sigma\omega } \delta_{\mu\rho } -3 \delta_{\eta\rho } \delta_{\eta\omega } \delta_{\mu\sigma }+\delta_{\mu\sigma } \delta_{\rho\omega }) +\delta_{\eta\mu } (-3 \delta_{\eta\sigma } \delta_{\eta\omega } \delta_{\nu\rho }+\delta_{\sigma\omega } \delta_{\nu\rho } -3 \delta_{\eta\rho } \delta_{\eta\omega } \delta_{\nu\sigma }+\delta_{\nu\sigma } \delta_{\rho\omega }) \big\}\big],
\notag 
}
where $\#$ is the number of $\eta$ contained in the Greek indices. 
} 
\beal{
\aver{\ddot\Ric_{\mu\nu}} 
=& -{2d\over (d-2) N}  g_{\mu\nu}, ~
\aver{\hat h_{\mu\nu} \dot R} 
= {8d \over N(d-2)} g_{\mu\nu}, ~ 
\aver{\ddot R }  
= {-2 d(1+d)(2+d) \over N(d-2) } .
}
Using these we obtain   
\beal{
\aver{\ddot G_{\mu\nu}}=& {d(d-1)(d+4) \over N(d-2)} g_{\mu\nu}. 
}
This result in fact is formally the same as the corresponding quantity computed in a free O($N$) vector model \cite{Aoki:2018dmc}. The intuitive reason for this is as follows. The AdS metric and the dS one are related by the analytic continuation of the AdS radial direction $z$ to the conformal time $z =i\eta$ (with an overall minus sign). The next-to-leading correction of the Einstein tensor has one warp factor, thus it gets one minus by the analytic continuation. This minus sign cancels the other one which comes from the one loop of the ghost fields. 
As a result, the next-to-leading corrections contribute to the VEV of the Einstein tensor with the opposite sign compared to the leading contribution:
\be 
\langle \hat G_{\mu\nu}\rangle =\frac{d(d-1)}{d-2}g_{\mu\nu}(-1 + \frac{d+4}{ N }). 
\ee

We interpret the expectation value of the Einstein tensor as the on-shell stress energy tensor in a dual bulk theory: $\langle \hat G_{\mu\nu}\rangle = T^{\rm bulk}_{\mu\nu}$. The holography is a duality such that there exists one-to-one correspondence between a normalizable solution of a bulk gravitational theory and a state in a dual boundary CFT \cite{Gubser:1998bc,Witten:1998qj,Balasubramanian:1998de}. This requires the on-shell stress-energy tensor corresponding to the CFT vacuum state to consist only of the cosmological constant term: $T^{\rm bulk}_{\mu\nu}= - \Lambda g_{\mu\nu}.$
Putting all together the cosmological constant and the de Sitter radius are determined up to the next-to-leading order as  
\beal{ 
\Lambda=\frac{d(d-1)}{d-2}(1 - \frac{d+4}{N }), \quad 
L_{\rm dS}^2={d-2\over2}(1 + \frac{d+4}{N }).  
}
In the higher spin/vector model correspondence, $1/N$ is identified with the Newton constant as $1/N\sim \hbar G_N /L^{d-1}_{\rm dS}$, and the $1/N$ expansion is interpreted as the loop expansion of the bulk fields in a dual gravity theory. Under the assumption of the correspondence between a free Sp($N$) vector model and a free higher spin theory with all even spins, the above result is identified with one loop corrections coming from vacuum bubble diagrams of the higher spin fields in the free higher spin theory on de Sitter spacetime. 

In conclusion, the holography by the flow equation method implies that the next-to-leading quantum corrections make de Sitter spacetime flatter and the cosmological constant smaller.%
\footnote{ 
Backreaction of a class of gravitational theories on constant curvature background has been very recently computed by the holographic renormalization group flow \cite{Ghosh:2020qsx}. It was found that the backreaction effect always decreases the bare cosmological constant in de Sitter spacetime, which is consist with the conclusion of this paper. The author would like to thank Elias Kiritsis for making him aware of the relevant reference.
}

\section{Discussion} 
\label{Discussion} 

We have investigated the dS/CFT correspondence using a free Sp($N$) vector model and a flow equation. 
To this end we have complexified the system and flowed the Sp($N$) ghost fields in parallel with the imaginary axis. As a result de Sitter spacetime has emerged with a flow parameter related to the conformal time in a suitable fashion. 
We have shown that this result is ensured by conformal symmetry and the holographic metric can still be interpreted as an information metric even in a case with a non-positive definite metric. 
We have also computed the vacuum expectation value of the Einstein tensor operator up to the next-to-leading order of the $1/N$ expansion. We have found that the $1/N$ corrections contribute to the cosmological constant in the opposite sign to the classical value, which suggests that quantum effects of gravity tend to flatten de Sitter spacetime. 
 
In section \ref{IM} we showed that the holographic metric to encode de Sitter spacetime can be regarded as an information metric even for a non-positive definite one. This non-positive definiteness is related to non-unitary nature of Sp($N$) vector models. However, it was pointed out that there exists a unitary operator to define an inner product which preserves under Hamiltonian dynamics of an Sp($N$) vector model \cite{LeClair:2007iy,Anninos:2011ui,Ng:2012xp}. It would be fascinating to explore the relation of this inner product to the information metric studied in this paper. (See also \cite{Geng:2019ruz						}.)

In section \ref{CC} we computed the one loop corrections to the cosmological constant in a free higher spin theory in de Sitter spacetime, which has the contribution with the opposite sign to the classical value. Implication of this result to quantum gravity on de Sitter spacetime is two-fold, which is based on a universal feature of the proposed dS/CFT correspondence. One is for its nature to incline for flatness. As long as a boundary CFT corresponding to gravity theory on de Sitter spacetime consists of ghost vector matter, the next-to-leading corrections to the cosmological constant seem to be negative in general by the argument given in this paper. It would be interesting to study whether this property holds including higher order corrections and explore the possibility for quantum effects to be an alternative solution of the flatness problem in the very early universe. 
The other is for its perturbative stability: A free higher spin gravitational theory on de Sitter spacetime is stable at least perturbatively. We suspect that this property holds in an interacting case as long as it is weakly coupled. However there is a conjecture that higher spin theories on de Sitter spacetime become unstable once non-perturbative corrections are taken into account \cite{Anninos:2012ft,Anninos:2013rza\if,Banks:2005bm\fi}. This conjecture has been made by computing a wave function of higher spin theories from Sp($N$) vector models on some compact manifolds and analyzing its divergent structure. It would be interesting to see such a non-perturbative instability from the flow equation approach. 

We argued that quantum corrections to the cosmological constant come from one loop vacuum bubble diagrams of higher spin fields, which are divergent in general. 
This suggests that a regularization scheme and a subtraction one in the bulk, which should be done in a covariant manner \cite{deHaro:2000vlm}, are automatically specified once a flow equation is fixed. This situation is very similar to that of a gauge fixing for diffeomorphism in the bulk \cite{Aoki:2019dim}. 
It is highly important to confirm the result presented in this paper by explicit computation of the one loop diagrams of higher spin fields. 

In this paper we mainly focused on a free Sp($N$) vector model. It is natural to extend this study to an interacting case. (See also \cite{Maldacena:2011jn,Maldacena:2012sf,Fei:2015kta}.) For example, one can introduce a double trace coupling to trigger the renormalization group flow. In three dimensions there exists the Wilson-Fischer fixed point, at which a theory achieves the conformal symmetry. It is conjectured that this CFT is dual to a free higher spin theory with the alternative boundary condition.  
Furthermore we can add the Chern-Simons interaction to gauge the Sp($N$) global symmetry as done in O($N$)/U($N$) vector models  \cite{Giombi:2011kc,Aharony:2011jz,Jain:2012qi}. This interaction produces a parity-violating effect, and a dual gravity theory is conjectured as an interacting higher spin gravity theory \cite{Vasiliev:1990en,Vasiliev:2003ev}.  
It would be interesting to test these conjectures from the flow equation approach and explore connection of different approaches of gravity on de Sitter spacetime \cite{Karch:2003em,Alishahiha:2004md,Kiritsis:2013gia,Das:2013mfa,Binetruy:2014zya,Narayan:2015vda} by incorporating techniques developed in this paper. 

We hope to give a new clue to solve the cosmological constant problem and to report progress on these issues in the near future. 

\section*{Acknowledgment}

The author would like to thank Sinya Aoki, Janos Balog, Tetsuya Onogi for helpful comments on the draft, and Tatsuma Nishioka for valuable discussion during East Asia Joint Workshop on Fields and Strings 2019 and 12th Taiwan String Theory Workshop. 
He is supported in part by the Grant-in-Aid of the Japanese Ministry of Education, Sciences and Technology, Sports and Culture (MEXT) for Scientific Research (No. JP19K03847).

\appendix 
\section{Derivation of \eqref{2ptflowchi}} 
\label{calculation}

In this appendix we give a brief derivation of \eqref{2ptflowchi}. This may be instructive to understand why in the flow equation approach operators are smeared not in the causal wedge but in all the imaginary direction. 

For this purpose we first prove a key equation such that the two point function of the flowed field is given by smearing the two point function of the original field:
\be 
\aver{\chi^a(\bm x_1;\xi_1) \chi^b(\bm x_2;\xi_2) } 
=\int_{\mbf R^d} d^d\check x_1' \rho(\check x_{12},\check x_1';\xi_1+\xi_2) \aver{\chi^a(\bm x_1') \chi^b(\bm x_2') }.
\label{flow2pt}
\ee
In what follows, we suppress the integration region for simple description if it is the total Euclidean space.
To show this, we first substitute \eqref{flowedop} into the left hand side.  
\beal{
\aver{\chi^a(\bm x_1;\xi_1) \chi^b(\bm x_2;\xi_2) } 
=&\Omega^{ab}\int {d^d \check x_1'} {d^d \check x_2'} e^{ - {(\check x_1-\check x_1')^2 \over 4\xi_1} } {1 \over (4\pi\xi_1)^{d\over2}} e^{ - {(\check x_2-\check x_2')^2 \over 4\xi_2} } {1 \over (4\pi\xi_2)^{d\over2}}  F_0(x_{12}+i\check x_{12}') , \notag
}
where we set $F_0(x)= {C \over ((x^k)^2)^{\Delta}}$ with $\Delta = {d-2\over2}$. We change the integration variables so that $\check x'^k_\pm={\check x_1^k \pm\check x_2^k \over \sqrt2}$. Then 
\beal{ 
\aver{\chi^a(\bm x_1;\xi_1) \chi^b(\bm x_2;\xi_2) } =& {\Omega^{ab} \over (4\pi\xi_1)^{d\over2}(4\pi\xi_2)^{d\over2}} \int {d^d \check x_+} {d^d \check x_-} e^{ - {(\check x_1- {\check x_+ + \check x_- \over \sqrt2})^2 \over 4\xi_1} - {(\check x_2-{\check x_+ - \check x_- \over \sqrt2})^2 \over 4\xi_2} } F_0(x_{12}+i\sqrt2\check x_{-})\notag.
}
We perform the integration in terms of $\check x_+'^k$ by gaussian:
\beal{
\aver{\chi^a(\bm x_1;\xi_1) \chi^b(\bm x_2;\xi_2) }=&{\Omega^{ab} \over (4\pi\xi_1)^{d\over2}(4\pi\xi_2)^{d\over2}}\({8\pi \xi_1 \xi_2 \over  (\xi_1+ \xi_2)}\)^{d\over2} \int {d^d \check x_-} e^{ -\frac{ \left(\check x_--\frac{\check x_1-\check x_2}{\sqrt2}\right)^2}{2(\xi_1+\xi_2)}  } F_0(x_{12}+i\sqrt2\check x_{-})\nn
=& {\Omega^{ab} \over (4 \pi(\xi_1+ \xi_2))^{d\over2}} \int { d^d (\sqrt2\check x_-) } e^{ -\frac{ \left(\sqrt2\check x_--\check x_{12}\right)^2}{4(\xi_1+\xi_2)}  } F_0(x_{12}+i\sqrt2\check x_{-}). \notag
}
This proves \eqref{flow2pt}.
Then let us compute 
\be 
I(x_{12},\check x_{12};\xi_+) 
:=\int d^d\check x_-' \rho(\check x_{12},\check x_-';\xi_+)F_0(x_{12}+i\check x_-') ,
\ee
where $\xi_+=\xi_1+\xi_2$. By using the Schwinger parameterization
\beal{ 
{1\over \lambda^s} = {1\over \Gamma(s)} \int_0^\infty d t  t^{s-1} e^{-\lambda t},
}
for $\Re s>0$, the integration can be written as 
\beal{
I(x_{12},\check x_{12};\xi_+) 
=& {C\over i^{2\Delta} (4\pi\xi_+)^{d/2}} \int d^d\check x' e^{- (\check x_{12} -\check x')^2\over 4\xi_+ } {1\over \Gamma(\Delta)} \int_0^\infty d t  t^{\Delta-1} e^{-(\check x'-ix_{12})^2 t} .  
}
For $\xi_+>0$, the integration in terms of $\check x'^k$ can be performed to give
\beal{
I(x_{12},\check x_{12};\xi_+) 
=& {C\over i^{2\Delta} (4\pi\xi_+)^{d/2}} {1\over \Gamma(\Delta)} \int_0^\infty d t  t^{\Delta-1} (\frac{\sqrt{4\pi\xi_+ }}{\sqrt{1+4\xi_+ t}})^d e^{-\frac{t (i x_{12} -\check x_{12})^2}{4 \xi_+  t+1}} .
} 
Changing the parameter by $4\xi_+ t={v \over 1-v}$, we find 
\beal{ 
I(x_{12},\check x_{12};\xi_+) 
=&{ 1 \over i^{2\Delta} (4\pi)^{{d\over2}} \xi_+^{\Delta}}\int_0^1 dv  v^{\Delta -1} e^{-{ (\check x_{12}-i x_{12} )^2\over 4\xi_+}v } . 
}
This derives \eqref{2ptflowchi}.

\bibliographystyle{utphys}
\bibliography{dSfromCFT}

\end{document}